\documentclass[aps,twocolumn,groupedaddress,footinbib,nobalancelastpage,prl]{revtex4-2}

\usepackage[colorlinks=true,citecolor=blue,linkcolor=red,urlcolor=red]{hyperref}
\usepackage{graphicx}
\usepackage{braket}
\usepackage[normalem]{ulem}
\usepackage{xcolor}
\usepackage{array}
\usepackage{orcidlink}
\usepackage{subcaption}
\usepackage{amsmath}
\usepackage{amssymb}
\usepackage{ragged2e}
\begin{document}

\title{Geometry-driven transitions in sparse long-range spin models with cold atoms}

\author{Alex Gunning\,\orcidlink{0009-0002-9028-3413}}
\author{Aydin Deger\,\orcidlink{0000-0002-6351-4768}}
\author{Sridevi Kuriyattil\,\orcidlink{0000-0002-9813-4714}}
\author{Andrew J. Daley\,\orcidlink{0000-0001-9005-7761}}

\affiliation{Department of Physics, Clarendon Laboratory, University of Oxford, Parks Road, Oxford OX1 3PU, United Kingdom}

\begin{abstract}

    We explore the influence of geometry in the critical behavior of sparse long-range spin models. We examine a model with interactions that can be continuously tuned to induce distinct changes in the metric, topology, and dimensionality of the coupling graph. This underlying geometry acts as the driver of criticality, with structural changes in the graph coinciding with and dictating the phase boundaries. We further discuss how this framework connects naturally to realizations in tweezer arrays with Rydberg excitations. In certain cases, the effective geometry can be incorporated in the layout of atoms in tweezers to realize phase transitions that preserve universal features, simplifying their implementation in near-term experiments.

\end{abstract}

\maketitle

Advances in atomic, molecular, and optical platforms have made it possible to engineer spin systems with controllable long-range interactions, enabling the realization of effective all-to-all \cite{britton_2012,Islam2013,zeiher_2017,vaidya_2018,hollerith_2022} and sparse \cite{Hung_2016,Bentsen_2019_sparse,qin_2019,Periwal_2021,Mivehvar_2021} coupling graphs. Deterministic sparse models have proven interesting for realization of fast scramblers--systems that saturate fundamental bounds on information spreading \cite{Susskind_1993,Hayden_2007,Barb_n_2013,belyansky_2020}--despite their limited graph connectivity \cite{bentsen_2019,Bentsen_2019_sparse,Hashizume_2021}. The fast scrambling regime corresponds to one limit in these models with rich many-body behavior, yet one in which the underlying geometry lacks a natural description of locality. To explore the geometrically meaningful regimes of these models, we allow continuous tuning of the relative interaction strength between spins at different distances. This tunability transforms the coupling geometry, unlocking two distinct limits of locality— linear and treelike \cite{Bentsen_2019_sparse, Periwal_2021}. In this work, we identify phase transitions in the low energy behavior of these models, driven by continuous changes in the geometry of the coupling graph. We show how the resulting phenomena can be probed experimentally, including how certain phase transitions can be realized directly with neutral atoms in tweezer arrays by mapping the geometry of the coupling graph to tweezer positions.

\begin{figure}[!t]
    \centering
    \includegraphics[width=1\linewidth]{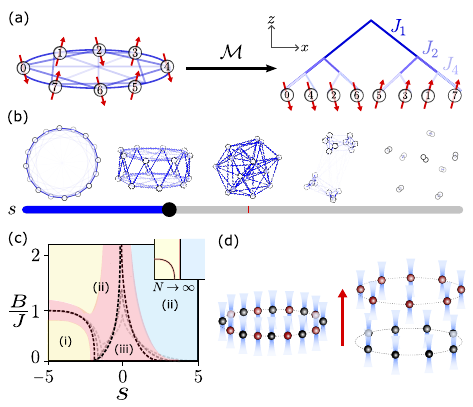}
    \caption{\small\justifying (a) 1D long-range Ising model on a power-of-two (PWR2) coupling graph, with transverse field $B$ in $x$-direction as in Eq. (\ref{eq:H}). A Monna map $\mathcal{M}$ transforms our 1D euclidean chain (left) into a hierarchal $2$-adic tree (right). (b) A slider schematic illustrating how continuous tuning of $s$ interpolates between different coupling geometries. (c) Schematic of quantum phase diagram for Eq. (\ref{eq:H}) as a function of geometry parameter $s$ and transverse field strength $B$. Dotted lines correspond to finite system sizes; solid lines indicate extrapolated thermodynamic behavior  (upper right box). Shaded regions denote phases characterized by area-law (yellow), volume-law (blue), and intermediate (red) entanglement entropy scaling. Phases (i) antiferromagnetic, (ii) paramagnetic, and (iii) geometric are labeled. (d) Rydberg atom array in a ring geometry. Odd sites (red) are vertically displaced, as indicated by the arrow, to mimic the tuning of the coupling parameter $s$ in (b) and yields the ``tambourine" geometry.}
    \label{fig:1}
\end{figure}

The model we consider introduces sparse coupling by constraining spins to interact only when their separation $d$ along a 1D ring is a power--of--two (PWR2). It features power law interactions $d^s$, where $s$ here allows the continuous tunability of the coupling graph connectivity. Recent work on this graph has revealed its
versatility: it reproduces key characteristics of dense all-to-all models--fast scrambling \cite{Bentsen_2019_sparse,bentsen_2019, Hashizume_2021}, complex dynamical transitions \cite{Kuriyattil_2023}, and metrologically useful entangled states \cite{Kuriyattil_2025}--while requiring only logarithmic connectivity per atom. Beyond emulation of all-to-all models, the tunable sparsity of the PWR2 graph provides a controlled route to exotic geometries. In one limit of $s$, couplings decay with distance, and a natural description is Euclidean [Fig.~\ref{fig:1}(a), left]. In the opposite limit, couplings grow with distance and a notion of locality is recovered by ordering sites by proximity along a tree [Fig.~\ref{fig:1}(a), right] \cite{barbon2012fast,Barb_n_2013,Gubser_2017,gubser_2018,gubser2019mixed,bentsen_2019}. In our analysis, we consider a complementary link to geometry which emerges when spins are recast as nodes on a weighted graph. Here, interaction strength is encoded in the edge weight: stronger couplings pull nodes closer together, while weaker couplings push them further apart. This representation reveals $s$ as a control knob for the topology and dimensionality of the interactions, evolving from ring--like through tambourine, sparse all--to--all and tree-like [Fig.~\ref{fig:1}(b), left $\rightarrow$ right]. 
We examine the resulting geometry-driven phase transitions [Fig.~\ref{fig:1}(c)], which offers insights on the interplay between geometry $s$ and quantum fluctuations $B$. Recent experimental realizations of related models in Rydberg arrays \cite{Bluvstein_2022,Graham2022,Bluvstein_2023,Evered2023,Xu_2024,Bernien2017} treat the graph’s geometry as distinct from the atom array. Instead, for regimes of $s$ where the weighted coupling graph has faithful representations in 3D, we show that this can be embedded directly into the physical layout of the Rydberg atoms [Fig.~\ref{fig:1}(d)], reproducing universal features of the relevant quantum phase transitions.

\paragraph{Power--of--two model.} The PWR2 graph can support various spin models; here we choose the transverse-field Ising case,  illustrated in Fig.~\ref{fig:1}(a), as its form closely mirrors the structure of the Rydberg Hamiltonian that we consider later. The Hamiltonian takes the form,
\begin{equation}\label{eq:H}
    H_{\rm PWR2} =  \sum_{i}^N\sum_{d \in \mathcal{D}}J_{d} \; S_i^zS_{i+d}^z\; + B \sum_i^N S_i^x ,
\end{equation}
where $B$ is the transverse field, and $J_d = J d ^s$ encodes the interaction strength between spins separated by distance $d \in \mathcal{D} = \left\{2^l \mid l =0,1,\dots,\log_2\frac{N}{2}\right \}$. The distance set $\mathcal{D}$ enforces a sparse hierarchy of couplings between spin pairs separated by powers of two \cite{bentsen_2019,Hashizume_2022,Kuriyattil_2023}. We consider a spin-$1/2$ model, where $S_{i}^{\alpha}=\sigma_{i}^{\alpha}/2$, with $\hbar = 1$ and $\alpha \in \{x,y,z\}$ with periodic boundary conditions imposed, $S_{i+d}^\alpha \equiv S_{(i+d \mod N)}^\alpha$. For $s\rightarrow -\infty$, the model reduces to the nearest-neighbor Ising model. As $s\to 0^-$, the dominance of nearest-neighbor bonds weaken and intermediate couplings become relevant. At $s=0$, all PWR2 bonds contribute equally, producing a sparse all-to-all model. For $s>0$, the dominant interaction shifts to the furthest pair $d=N/2$, giving an ultrametric structure that maps to a binary tree defined by the $2$-adic distance. The Monna map $\mathcal{M}$ \cite{MONNA1952} [Fig.~\ref{fig:1}(a)] exposes this structure by reversing binary site indices. In the $s\to \infty$ limit, the model becomes the furthest-neighbor Ising model.

\paragraph{Phase diagram.}  We begin by analyzing the classical limit of the PWR2 model, by letting the transverse field $B/J \to 0$ in Eq.~(\ref{eq:H}).
We also set $J<0$ to impose that as $s\to-\infty$ we have the \textit{antiferromagnetic} (AFM) ground-state. A renormalization factor $(2/N)^s$ is introduced in the $J_d$ term if $s>0$ to constrain interaction strength to a maximum value $J$. As our model has no known exact solution, we leverage our knowledge of the underlying geometry of the system to locate the critical points. The coupling graph undergoes distinct structural transformations as $s$ is tuned, which are captured by the local intrinsic dimensionality of our weighted coupling graph in Fig.~\ref{fig:1}(b). This diagnostic measures how effectively the graph structure can be embedded in a lower-dimensional space without significant loss of information \cite{LID, 6406405,levina2004maximum}. It is built from our $N \times N$ coupling matrix, rather than the full $2^N$ Hamiltonian, so we can detect critical points at much larger $N$ (see Supplemental Material). The analysis indicates phase boundaries at $s=-2$, $s=0$ and some $s\gg 0$, serving as a heuristic for the possible critical points in the thermodynamic limit, and guiding our energy gap analysis below.

Phase transitions can be characterized by critical values of the control parameter where the energy gap $\Delta E$ closes in the thermodynamic limit $N\rightarrow \infty$. We define the energy gap $\Delta E$ as the minimum energy difference between the ground-state manifold  and the first excited-state manifold ($1$ES),
\begin{equation}
    {\Delta E} = \min_{i \in 1 \rm{ES}}(E_i - E_0),
\end{equation}
where $E_0$ is the ground-state energy and the minimization excludes all states that are degenerate with the ground-state, to account for symmetries and spectral redundancies inherent to the PWR2 interaction graph. In each geometrical regime, we identify a candidate ground-state and construct the lowest-energy excitation that is consistent with minimal disruption of the dominant bond (nearest-neighbor for $s<0$ and furthest-neighbor for $s>0$). With this we calculate the energy gap between the manifolds, and track it as a function of $s$ to predict critical points where the gap closes. Our analytical predictions are corroborated by exact numerical diagonalization for small system sizes and further supported by Monte Carlo methods for larger systems \cite{krauth_2006,robert_MCMC_2011} (see Supplemental Material).

\begin{figure}[t]
    \centering
    \includegraphics[width=1\linewidth]{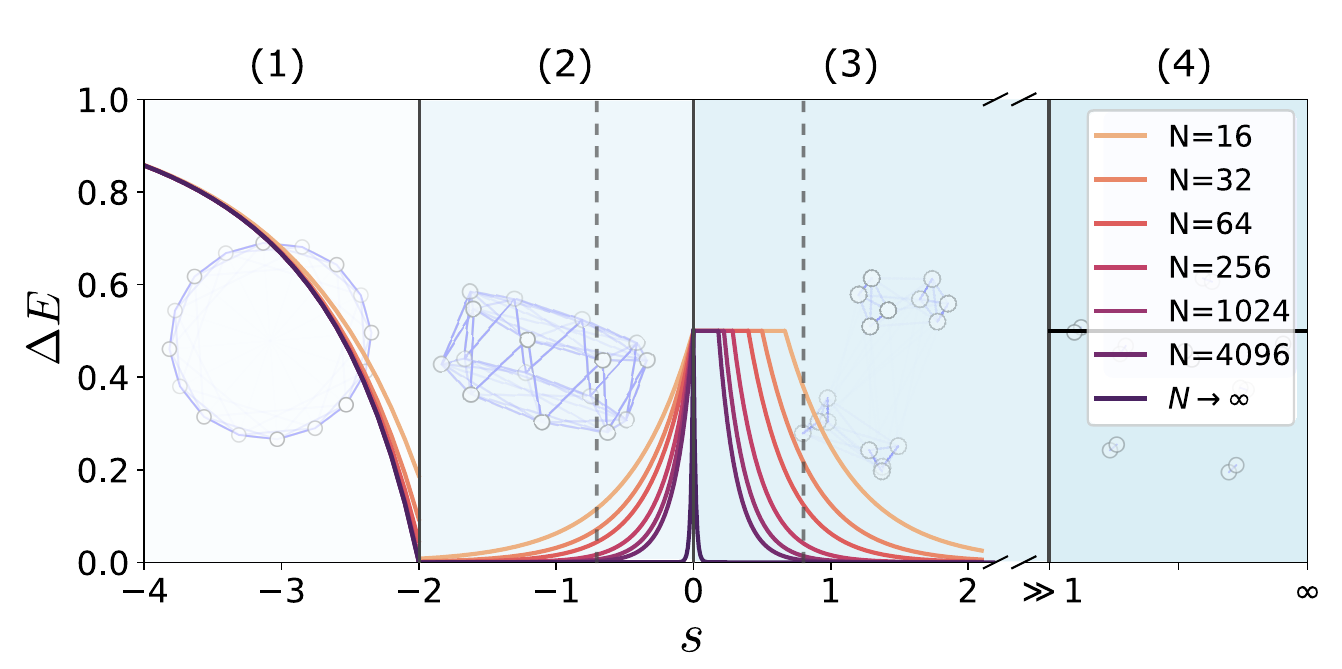}
    \caption{\small\justifying Analytical energy gap $\Delta E$ as a function of coupling parameter $s$ for classical limit of PWR2 model $B/J=0$ in Eq.~(\ref{eq:H}). Four distinct phases (labeled (1)-(4)) are present in the infinite system, separated by critical points (solid lines). At finite sizes an additional gap opens around $s=0$, giving rise to an extra region (dashed lines). Semi-transparent overlays (blue) depict the effective geometry associated with each region, illustrating how changes in connectivity coincide with the phase boundaries.}
    \label{fig:2}
\end{figure}

The heuristic guidance of the local intrinsic dimensionality analysis suggests four distinct phases of the PWR2 model as $s$ is tuned, and our gap analysis confirms this structure [Fig.~\ref{fig:2}]. For $s<-2$, the coupling graph is ring-like and dominated by nearest-neighbor bonds, yielding an antiferromagnetic ground-state with a finite energy gap. The gap closes at $s=-2$ in the thermodynamic limit, precisely where the intrinsic dimensionality departs from $1D$. Entering the regime $-2<s<0$, the system becomes highly frustrated and gapless: intermediate interactions compete and the effective graph dimension rises continuously. At $s=0$, the system realizes the recursive PWR2 ground-state (see Supplemental Material), which optimally satisfies all power-of-two bonds and reflects the all-to-all $\log_2(N)$-dimensional structure of the underlying graph. The gap is strictly open only at this single point in the infinite system, but at finite $N$ it broadens into a peak (captured within the dotted lines of Fig.~\ref{fig:2}). Convergence to the thermodynamic limit here is extremely slow, with the gap closing fully only at very large system sizes ($N>10^{10}$). Its robustness at experimentally accessible scales is therefore of particular interest, as this finite-size region is expected to persist in all practical settings. For $s>0$, the recursive PWR2 ground-state remains optimal, but an extensive manifold of $2^{N/2}$ excited states collapses onto it, each consistent with the treelike antiferromagnetic order dictated by the $2$-adic geometry, culminating in the Monna-mapped AFM of Fig.~\ref{fig:1}(a) (right). At this stage the gap reopens, marking the final phase; the corresponding critical point drifts to $s\to\infty$ as $N\to\infty$, in full agreement with the intrinsic dimensionality framework. Analytical derivations of the thermodynamic limits of the critical points, along with additional details of the corresponding ground-states, are provided in the Supplemental Material.
Having established the criticality induced by tuning the geometry of the PWR2 model, we now consider the quantum case ($B/J\neq0$) to probe the robustness of these phases in near-term experimental realizations, where quantum fluctuations are unavoidable. This analysis also lays the groundwork for comparison with the Rydberg model, in which fluctuations act as a natural probe of universal behaviour. We employ exact diagonalization for small system sizes $N\leq16$ and density matrix renormalization group (DMRG) methods with matrix product states (MPS) for larger $N$ \cite{WhiteFeiguin2004, DaleySchollwoeck2004, Schollwoeck2011, FishmanStoudenmire2022, ITensor2022}. For $s > 0$, we exploit our alternative notion of locality by applying the Monna map to the MPS site ordering. This reorders sites according to their $2$-adic representation, placing strongly correlated (distant) sites adjacent in the MPS structure. As a result, long-range correlations are encoded locally,
reducing the reliance on long bonds and improving convergence. To identify and characterize the critical points, we turn to the bipartite entanglement entropy, which serves as a sensitive probe for quantum phase transitions.

The phase diagram in Fig.~\ref{fig:1}(c) illustrates the interplay between the geometry parameter $s$ and transverse field $B$. Phases are distinguished by entanglement entropy scaling: area law (yellow), volume law (blue), and intermediate scaling (red). For small $B$, the quantum phase diagram can be directly understood from the classical phase diagram and energy gap analysis, as the transverse field introduces spin flips into the ground-state. The cost of these flips---and thus the strength of $B$ required to drive a transition to the paramagnet---is determined by their position in the classical excitation spectrum and whether that spectrum is \textit{gapped}. In the nearest-neighbor AFM limit ($s \to -\infty$), there is a well-known transition at $B=J$ from the AFM phase (i) to the paramagnet (ii), both showing area-law scaling, with a critical conformal field theory (CFT) scaling region (red) collapsing to the transition point as $N \rightarrow \infty$ \cite{Callan_1994,vidal2003,Pasquale_Calabrese_2004,Carr2009}. In \textit{Classical Phase 1}, as $s$ moves towards $s_c=-2$, the energy gap closes, lowering the $B$ needed to destabilize order. At $s=-2$ the system is gapless, so any finite $B$ drives it directly into an area-law paramagnet (although a (red) intermediate scaling region exists at finite sizes). In \textit{Phase 2}, the gap protects the classical phase (iii) at moderate $B$, but vanishes in the thermodynamic limit, collapsing the phase to a critical line at $s=0$ with critical scaling (red). The high graph connectivity stabilizes this line against large $B$ [Fig.~\ref{fig:1}(a), inset]. The transverse field merges \textit{Classical Phase 3 \& 4}, into a single quantum critical line. This stems from the fact that between these boundaries there was never a distinct ordered phase, just a collapsing manifold of $2^{N/2}$ states. The quantum critical line originates near the onset of manifold formation (dotted line in Fig.~\ref{fig:2} that collapses onto $s=0$ in thermodynamic limit) as that’s where the first true level crossing and change of the ground-state character occur. The (blue) volume-law paramagnet dominates for all $B$ beyond this phase line. In this regime the entanglement exhibits area-law scaling when measured in the Monna basis, indicating the observed volume-law is an artifact of our choice of measurement space (Euclidean rather than the appropriate ultrametric geometry). Representative slices of the half-chain entanglement entropy highlight these distinct behaviors across the quantum phase diagram (see Supplemental Material).

\paragraph{Experimental implementation.} Related long-range interacting models have been successfully implemented in cavity-QED platforms \cite{Hung_2016,bentsen_2019,Bentsen_2019_sparse,qin_2019,Periwal_2021} and in Rydberg tweezer arrays \cite{jaksch_2000,Bluvstein_2022,Graham2022,Bluvstein_2023,Evered2023,Xu_2024,Bernien2017}. Cavity-QED realizations typically employ collective large-spin modes and spin-exchange interactions, while Ising interactions remain non-trivial. Rydberg setups instead exploit Trotterization and stroboscopic schemes to engineer effective long-range couplings \cite{Nill_stroboscopic2025,Zhao2023FloquetRydberg,koyluoglu2024floquetengineeringinteractionsentanglement,Kuriyattil_2025,Bluvstein_2023,Bluvstein_2022}. The scheme we propose offers a simple, complementary path to accessing specific PWR2 ground-state transitions in near-term experiments using neutral atoms in tweezer arrays.

We consider the Rydberg Hamiltonian \cite{Saffman_2010_Quantum_info_ryd, Adams_2020, Henriet_2020, Morgado_2021} in a spin-$1/2$ basis,
\begin{equation}\label{eq:H_vdw}
H_{\rm Ryd} \;=\; \sum_{i\neq j}^N V(r_{ij})\, S_i^z S_j^z \;+\; B\sum_{i}^N S_i^x \;+\;\Delta\sum_i^NS_i^z,
\end{equation}
where $V_{ij} = \frac{C_6}{r_{ij}^6}$ denotes the van der Waals interaction between atoms separated by a real-space distance $r_{ij} \equiv \|\mathbf{r}_i-\mathbf{r}_j\|$, and $\{\mathbf{r}_i\}$ are the physical 3D positions of the atoms. $\Delta$ is the laser detuning and we consider the resonant-driving regime ($\Delta = 0$). In order to reproduce the phase transitions of the PWR2 model that we introduced above, we need an analogous \textit{continuous} control parameter. Since the van der Waal term fixes the interaction exponent at $s=-6$, a direct tuning of $s$ is not available. Our solution is to directly embed the emergent interaction graph of the PWR2 model into the physical layout of the atoms in the Rydberg Hamiltonian. As $s \to 0$, the effective dimensionality of the PWR2 coupling graph grows as $\log_2(N)$, extending beyond what can be naturally embedded in three physical dimensions. Within a Rydberg implementation, this restricts direct real-space realizations to our phase line stemming from $s_c=-2$ in Fig.~\ref{fig:1}(c). For our proposal, the atoms in the tweezer array are initially arranged in a ring geometry in the $x$--$y$ plane. Atoms at every odd site are then lifted out of the plane by a vertical displacement $h$, such that $z_{i} = 0$ ($i$ even) and $z_i=h$ ($i$ odd) in the atom positioning $\mathbf{r}_i = (x_i,y_i,z_i)$. This forms the tambourine geometry shown in Fig.~\ref{fig:1}(b), with $h$ as the control parameter. We refer to this constrained realization as the \textit{tambourine Rydberg model}. 

\begin{figure}[t]
    \centering
    \includegraphics[width=1.0\linewidth]{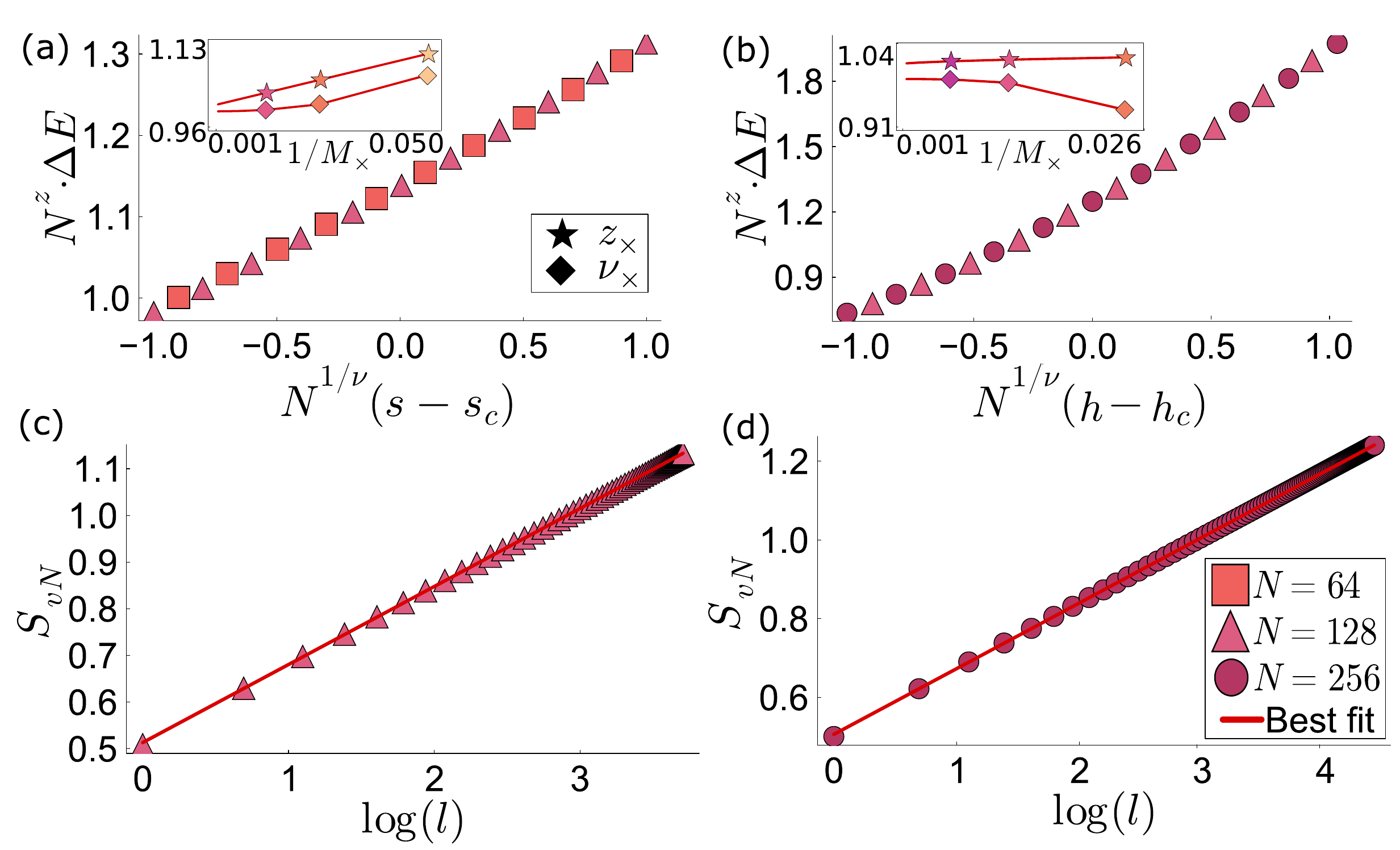}
    \caption{\small\justifying Finite-size scaling analysis of the energy gap $\Delta E$ and half-chain entanglement entropy $S_{vN}$ for the PWR2 and tambourine Rydberg model showing consistency with 2D Ising class. (a) Scaling collapse of the rescaled gap $N^z \Delta E$ for the PWR2 model at system sizes $\times=(64,128)$, optimized at $\nu_{\times} \approx 0.993$ and $z_{\times}\approx1.026$ at critical point $s_{c\times} \approx -3.3939 $ for $B/J=0.8$. Inset: Thermodynamic limit $z_{\infty}=1.000(2)$ and $\nu_{\infty}=0.99(1)$ from the best fit convergence of $z_{\times}$ and $\nu_{\times}$ with $M_{\times}=\sqrt{2}N$. (b) Extraction of the central charge from the entanglement entropy [Eq.~\ref{eq:c}] at $N=128$, with best linear fit yielding $c=0.5000(3)$ at the thermodynamic value of $s_c$. (c) Scaling collapse for the tambourine Rydberg model at system sizes $\times=(128,256)$, optimized at $\nu_{\times} \approx 0.98$  and $z_{\times} \approx  1.01 $ at $h_{c\times} \approx 0.9935$ for $B/J=0.8$. Inset: Thermodynamic limit $z_{\infty}=1.01(5)$ and $\nu_{\infty}=0.99(4)$ from the best fit convergence of $z_{\times}$ and $\nu_{\times}$. (d) Extraction of the central charge at $N=256$, with optimized linear fit when $c=0.5000(9)$.}
    \label{fig:crit}
\end{figure}

\paragraph{Critical behavior of the Power--of--two and tambourine Rydberg model.} To determine whether the phase transition induced by rearranging the atoms in our Rydberg model reproduces the transition obtained by tuning $s$, we examine whether both models belong to the same universality class at criticality. We first establish the universality class of the phase line in the PWR2 model. We use pairwise finite-size scaling based on the structure-factor $\xi_N(s)$ crossings described in the Supplemental Material. The size-independent crossing of $\Phi_N(s)=\xi_N(s)/N$ and $\Phi_{2N}(s)$ defines the pairwise estimate $s_{c\times}$ at $\times = (N,2N)$. We extract $\nu_{\times}$ from the ratio of derivatives $\frac{d\phi_{2N}}{ds}/\frac{d\phi_N}{ds}$ and $z_{\times}$ from the ratio of energy gaps $\Delta E_{2N}/\Delta E_N$ at the crossing \cite{fisher1972,Fendley_2004,Sandvik_2010,buyskikh_2019}. Using these optimal exponents, the rescaled gaps $N^z\Delta E$ collapse onto a single universal curve, as shown in the main panel of Fig.~\ref{fig:crit}(a). The inset of Fig.~\ref{fig:crit}(a) shows the weighted least-squares fits of $\nu_{\times}$ and $z_{\times}$ versus $M_{\times}=\sqrt{N(2N)}=\sqrt{2}N$, yielding the thermodynamic values $\nu_\infty=0.99(1)$ and $z_\infty=1.000(2)$, fully consistent with the 2D Ising class.
If the system is conformally invariant at $s=s_c$, additional universal information can be extracted using CFT \cite{Pasquale_Calabrese_2004,Cardy_2010}. In particular, for a 1D chain with periodic boundary conditions, the bipartite entanglement entropy scales as \cite{Callan_1994,vidal2003,Cardy_2010},  
\begin{equation}\label{eq:c}
    S_{vN}(L) =  \frac{c}{3} \, \ln\!\left( \frac{L}{\pi } \, \sin\!\frac{\pi l}{L} \right) + \mathcal{O}(1),
\end{equation}
where $c$ is the central charge of the underlying CFT and $\mathcal{O}(1)$ denotes non-universal constant corrections. In Fig.~\ref{fig:crit}(b) it is possible to extract a value of $c = 0.5000(3)$ for the PWR2 model at $s_c$. All of the extracted exponents are clear indicators that the phase line in Fig.~\ref{fig:1}(a) belong to the 2D Ising class.

We now examine whether $h_c$ of the tambourine Rydberg model belongs to the 2D Ising universality class. In Fig.~\ref{fig:crit}(c) we extract similar critical exponents for thermodynamic values of $z_{\infty}=1.01(5)$ and $\nu_{\infty} = 0.99(4)$ at a critical point $h_c = 0.993(3)$. In Fig.~\ref{fig:crit}(d) we also extract a value for the central charge $c = 0.5000(9)$. All of these extracted exponents capture the 2D Ising class. The analysis here demonstrates that a phase transition with the same critical behavior as the PWR2 model can be induced in the tambourine Rydberg Hamiltonian by tuning the atomic arrangement. A mapping $s(h)$ can be obtained by optimizing the fit of $J_{i,j}(s)\propto |i-j|^{\,s}$ to $V_{i,j}(h)\propto r_{i,j}(h)^{-6}$ in the classical limit (see Supplemental material). This mapping shows that the Rydberg critical point $s(h_c)$ is shifted from $s_c$ due to non--PWR2 bonds. A dual--species scheme \cite{Kevin_Singh_2022,Kevin_Singh_2023,Beterov_2015, Samboy_2017,Ireland_Paul_2024} that selectively suppresses these couplings restores the correct critical behavior (e.g. $s(h_c) = -1.998(7)$ which closely matches with $s_c = -2.0$) and also captures the PWR2 ground-state beyond the transition.

\paragraph{Conclusion.} 

The PWR2 model provides an experimentally accessible setting in which geometry itself dictates quantum phases through continuous tuning of the coupling graph. Our tambourine Rydberg model allows realization of some of these phase transitions, without Trotterization or Floquet schemes, especially when simplified using dual-species encoding. A natural extension of this work is to introduce disorder and assess whether geometric connectivity in these regimes counteracts localization and sustains scrambling \cite{anderson_1958,Basko_2006}. The frustrated gapless regimes we identify also motivates exploring possible connections to spin-liquid-like behavior, a longstanding challenge in strongly correlated quantum systems \cite{anderson_1973,ramirez_1994,lee_2008,balents_2010}. More broadly, the geometric framework developed here applies to a wide class of interaction graphs beyond power-of-two connectivity, opening the door to experiments and theory that probe and classify quantum many-body behavior by actively reshaping the geometry of interaction graphs.\\

In compliance with EPSRC's open access initiative, the data in this paper will be available from \cite{data_open_access}.\\

\begin{acknowledgments}
 We thank Jonathan Pritchard, Gregory Bentsen, Chris Hooley, Tomohiro Hashizume, Monika Schleier-Smith, Sebastian Schmid, and Ian B. Spielman for helpful discussions. This work was supported by the EPSRC through the QQQS programme grant (EP/Y01510X/1) and by the QCi3 hub (EP/Z53318X/1).

\end{acknowledgments}

\bibliography{bibliography.bib}

\widetext

\begin{center}
\vskip0.5cm
{\Large Supplemental Material}
\end{center}
\vskip0.4cm

\setcounter{section}{0}
\setcounter{equation}{0}
\setcounter{figure}{0}
\setcounter{table}{0}
\setcounter{page}{1}
\renewcommand{\theequation}{S\arabic{equation}}
\renewcommand{\thefigure}{S\arabic{figure}}
\renewcommand{\thesection}{S\arabic{section}}
\renewcommand{\thesection}{S\arabic{section}}

\section{Local intrinsic dimensionality analysis}

\paragraph{Local Intrinsic Dimensionality.} The representation of data in high-dimensional spaces can often be reduced to a lower-dimensional manifold without significant information loss. The intrinsic dimension of a data set is the lowest dimensional space in which the data can be effectively embedded. Local intrinsic dimensionality provides a way to probe this property by examining only the $k$ nearest neighbours of each data point, under the assumption that the neighbourhood structure reflects the underlying dimension \cite{LID, 6406405}. In this study, we estimate local intrinsic dimensionality using a maximum likelihood estimator following the approach in \cite{levina2004maximum}. 

\paragraph{Maximum Likelihood Estimator.} For each data point, the distances to its $k$ nearest neighbors were computed. If we denote these ordered distances as $d_1, d_2, \dots, d_k$, the estimator takes the form,
\begin{equation}
\widehat{\mathrm{LID}} = \left( \frac{1}{k-1} \sum_{i=1}^{k-1} \log \frac{d_k}{d_i} \right)^{-1}.
\end{equation}
 By looking at the relative distances to neighbours, we're capturing how densely or sparsely the data fills the space, which directly reflects its effective dimension. We repeated this calculation for every point in the graph and average it to obtain a global estimate of the intrinsic dimension. 
 \begin{figure}[h!]
    \centering
    \includegraphics[width=0.5\linewidth]{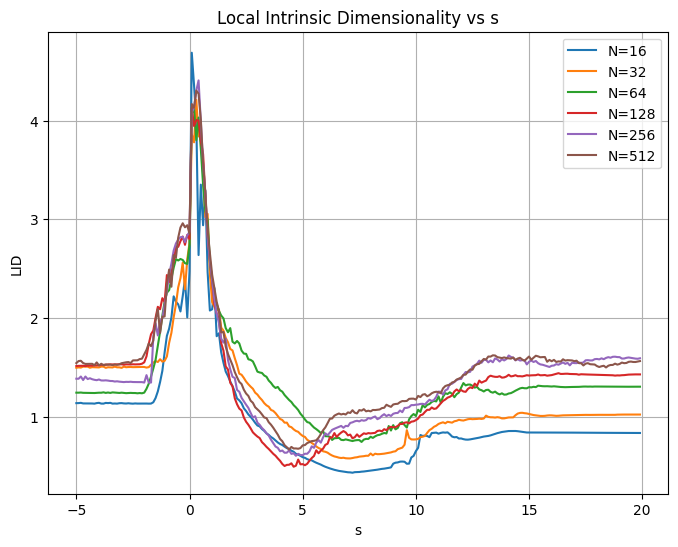}
    \caption{Local intrinsic dimensionality as a function of the coupling parameter $s$ for increasing system sizes, taking $k=\log_2(N)$. The data shows stable 1D behavior until $s=-2$, where a sharp peak emerges, followed by a growth in dimensionality up to $s=0$. Beyond $s=0$, the dimensionality decreases and settles into a plateau at $s>>0$. A finite-size dip appears at small positive $s$, but shifts back toward $s=0$ as the system grows, consistent with the classical phase diagram.}
    \label{fig:LID}
\end{figure}
 \paragraph{Numerical results.} When applying this method to the weighted coupling graphs of Fig.~\ref{fig:1}(b), we observe that the estimated intrinsic dimensionality changes sharply at critical values of the coupling parameter as shown in Fig.~\ref{fig:LID}. These shifts coincide with the established critical points of the model. For $s\ll0$, the dimensionality remains low and constant, reflecting the underlying one-dimensional ring structure. At $s=-2$, the dimensionality jumps abruptly, signaling the emergence of the tambourine geometry with higher dimensionality. The local dimension continues to increase as we approach $s=0$ where it is at a maximum. For $s>0$ the dimensionality dips again. At intermediate $s$ there is a finite size effect that dips, but is shown to push back toward $s=0$ with increasing system sizes. At some $s>>0$ the dimensionality plateaus again reaching the final region of our model. The point at which the system plateaus pushes back with increasing system size. The final critical points this analysis indicates are therefore $s=-2, s=0$ and $s \gg0$. It is striking that such a simple analysis of the $N \times N$ coupling matrix, can reveal clear qualitative transformations at phase boundaries without requiring the heavy machinery of full exact diagonalization and spectral analysis.

\section{Markov chain Monte carlo (MCMC) analysis}

As the model is classical, we can simulate much larger system sizes than are accessible to exact diagonalization by using Markov-chain Monte Carlo \cite{krauth_2006,robert_MCMC_2011}. This approach is far more efficient but comes with the trade-off that spin configurations are sampled stochastically rather than exhaustively. As a result, accessing the full low-energy manifold is more difficult, and statistical fluctuations inevitably appear in the data. These can be reduced, though never entirely removed, by increasing the number of samples. Close to the critical value $s_c$, critical slowing down further reduces accuracy because spin configurations become correlated over long length scales. 

\paragraph{Metropolis Algorithm.} The algorithm works by setting an initial random configuration of up and down spins. An attempt is made to flip each spin in the chain sequentially, if a flip is energetically favorable it is accepted and if a flip is unfavorable it is rejected unless a randomly generated number $c \in [0,1]$ is less than the Boltzmann weighting $e^{-\beta \Delta E}$ associated with the spin flip. As we are interested in the low lying spectrum (ground-state and first excited state), we set the temperature low, or correspondingly set $\beta$ high. 
\begin{figure}
    \centering
    \includegraphics[width=0.4\linewidth]{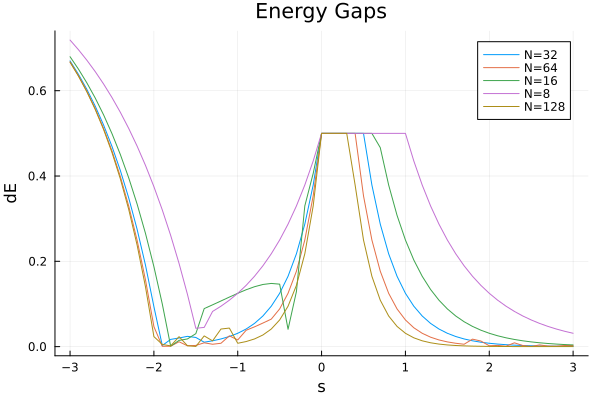}
    \caption{Numerically computed energy gaps using the Monte Carlo Markov Chain method, which shows good agreement with the analytical results of Fig.~\ref{fig:2}, except for numerical errors and finite-size effects that arise in the simulation in the gapless regimes.}
    \label{fig:mcmc}
\end{figure}

\paragraph{Numerical results.} In Fig.~\ref{fig:mcmc}, there is good agreement between the numerical results and the analytical results of Fig.~\ref{fig:2}. The MCMC struggles in the gapless regions as the algorithm encounters numerous nearly degenerate states with similar energies, which prevents convergence to the true ground-state. This is improved with increased sampling sizes but this quickly becomes inefficient with growing system sizes. Our analytical method below is the optimal method to properly scale to the thermodynamic limit. 

\section{Analytical ground-state analysis}

\paragraph{Classical Phase 1.} The coupling graph exhibits a ring-like structure dominated by nearest-neighbor interactions, naturally suggesting an AFM ground-state. The first excited state consists of two domain walls (DW), minimizing the disruption to nearest-neighbor bonds. The energy gap can be naturally extended to the thermodynamic limit, as both states admit well-defined configurations for any $N$. The corresponding energy gap is computed as $\Delta E = E_{\rm 2 \rm{DW}} - E_{\rm AFM}$ using the explicit form of $H$. The critical point marking the transition out of AFM order $s_1\propto \frac{1}{N}-2$ approaches $s=-2$ as $N\rightarrow \infty$, matching perfectly with the transition out of our ring topology in our intrinsic dimensionality analysis. 

\paragraph{Classical Phase 2.} A tambourine-like structure emerges in our coupling geometry. This picture maps to a gapless phase, comprising a manifold of states characterized by domain wall counts that reflect the growing strength of intermediate interactions. As $s\rightarrow 0^-$ the tambourine distorts as intermediate-range bonds gain strength, and a transition occurs into a region characterized by the recursive PWR2 ground-state. This ground-state attempts to optimally satisfy all PWR2 bonds, and has a recursive construction.  For a system size $N=2^l$, the ground-state is generated by taking the previous $N=2^{l-1}$ ground-state that optimizes the $l-1$ PWR2 couplings, and appending the inverse to it, ensuring satisfaction of the extra furthest-neighbor coupling introduced,
\begin{align}
    \ket{N = 2} &= \ket{\uparrow \downarrow}\\
    \ket{N = 4} &= \ket{\uparrow \downarrow \downarrow \uparrow} \\
    \ket{N = 8} &= \ket{\uparrow \downarrow \downarrow \uparrow \downarrow \uparrow \uparrow \downarrow} \\
    \ket{N = 16} &= \ket{\uparrow \downarrow \downarrow \uparrow \downarrow \uparrow \uparrow \downarrow \downarrow \uparrow \uparrow \downarrow \uparrow \downarrow \downarrow \uparrow}\\
    \vdots & \nonumber 
\end{align}
The first excited state is a single spin flip that breaks only one furthest-neighbor bond. The formula for the energy difference of a single spin flip at site $j$,
\begin{equation}
     \Delta E = -\frac{1}{2}\sum_i^N J |i-j|^{s}S_i S_j,
\end{equation}
gives rise to an analytical formula for the energy gap in this region as a function of $N$, $\Delta E (N) = \frac{1}{2} J |\frac{N}{2}|^{s}$. As $N\rightarrow \infty$ the critical point where the energy gap opens $s_2 \propto 1/\log(\frac{N}{2}) \rightarrow 0$. At $s=0$, our formula shows the energy gap becomes independent of $N$, $\Delta E = \frac{1}{2}$ as shown in Fig.~\ref{fig:2}. Here, the coupling structure takes on an all-to-all graph constrained to PWR2 couplings. 

\paragraph{Classical Phase 3.} For $s>0$, the furthest-neighbor bond replaces the nearest-neighbor bond as the dominant coupling. A treelike hierarchal structure of the PWR2 geometry is apparent in the graph structure, with clusters of spins forming dependent on their position in the $2$-adic tree. The point at which the flat energy gap plateau sharply starts to collapse, marks the onset of furthest-neighbor bond dominance. Here, the first excited state consists of two spin flips on top of the recursive PWR2 ground-state. Specifically, flipping spins at sites $i$ and $i+N/2$ is energetically favored over a single spin flip, as this configuration avoids disruption of any furthest-neighbor bonds. The value of $s$ at which the gap begins to close occurs when $2(\frac{2}{N})^{s} = \frac{1}{2}$. In the thermodynamic limit this point $s \propto 1/\log(\frac{2}{N}) \rightarrow 0$. As $s$ increase from here and the influence of short range bonds continues to diminish, the first excited manifold becomes degenerate with the ground-state. This region is resultantly gapless as the exponential number ($2^{N/2}$) of states that satisfy furthest-neighbor bonds begin to collapse onto the ground-state. The analytical form of the energy gap in this region has the form $\Delta E(N) = 2J(2/N)^{s}$. As $N\rightarrow \infty$ the critical point where the energy gap closes satisfies $s\rightarrow 0$. As shown in Fig.~\ref{fig:2}, the thermodynamic scaling of all these points compress the energy gap plot into a sharp, isolated peak at $s=0$. 

\paragraph{Classical Phase 4.} This marks the point where all the furthest-neighbor states have collapsed into the ground-state manifold, and a gap reopens. The last state to enter this manifold, marking the transition, is crucially the treelike AFM state, $\ket{\uparrow\uparrow\dots\uparrow\uparrow\downarrow \downarrow\dots\downarrow\downarrow}$, found by Monna mapping the Euclidean AFM. The transition into this region is pushed to $s\rightarrow \infty$ in an infinite system, rendering it inaccessible for any finite coupling $s$.\\

\section{Entanglement entropy cuts of quantum phase diagram}

\begin{figure}
    \centering
    \includegraphics[width=0.35\linewidth]{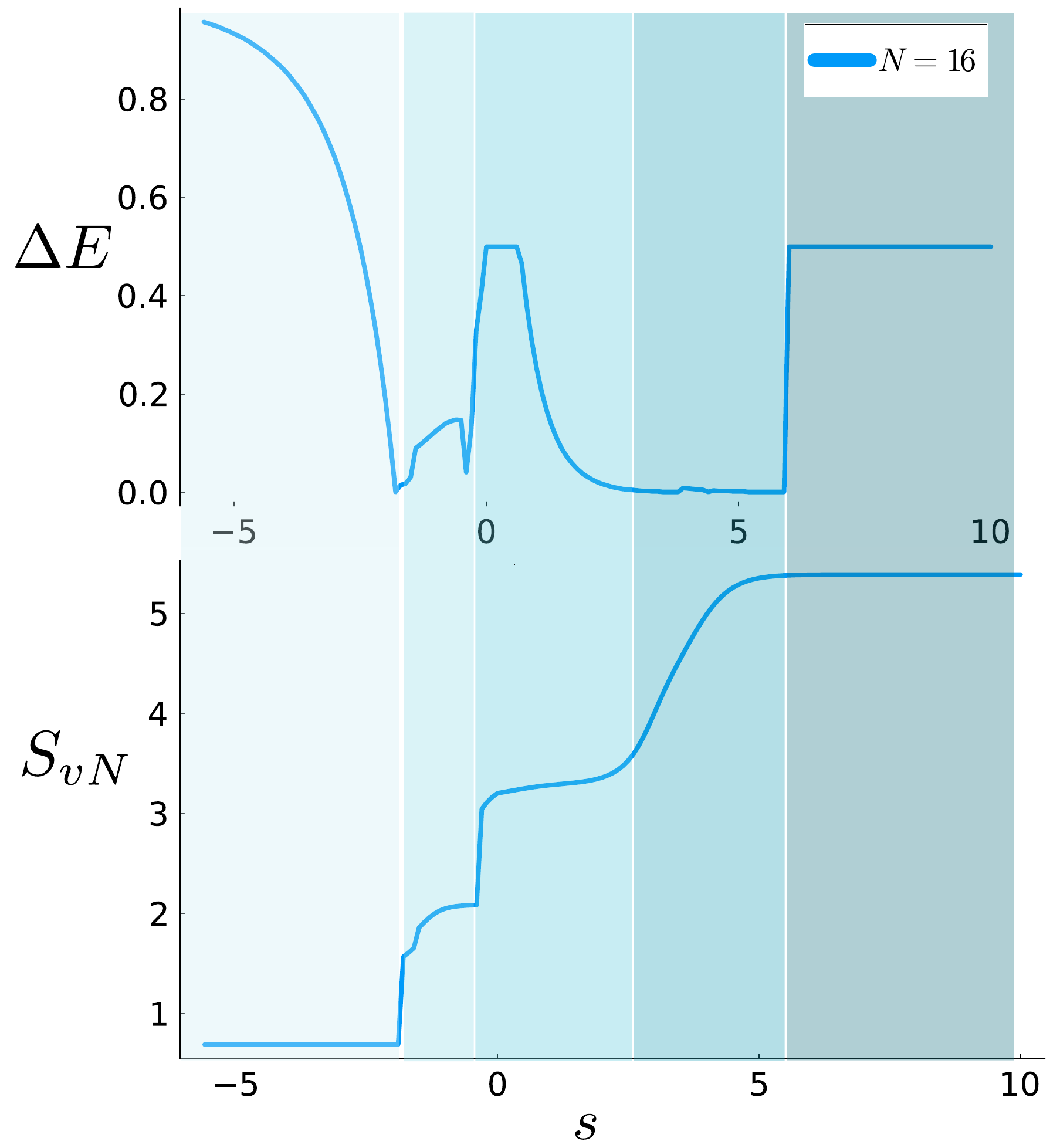}
    \caption{Bipartite entanglement entropy $S_{vN}$ at weak transverse field capturing surviving behavior of classical phases. Top: Energy gap analysis $\Delta E(s)$ for $N=16$, whose boundaries at finite sizes account for the features in $S_{\mathrm{vN}}$. Bottom: von Neumann entanglement entropy $S_{\mathrm{vN}}$ for $N=16$ at $B/J=0.1$ versus the geometry parameter $s$, with shaded bands marking regimes inherited from the $B{=}0$ phase diagram. Together these panels show that the finite-size structure of the classical phases survives at small transverse field. Lower figure reveals the connection between these regions and the classical phase diagram with matching colors representing the origin of each behvaior.}
    \label{fig:lowB}
\end{figure}

\begin{figure}[b]
    \centering
    \includegraphics[width=0.6\linewidth]{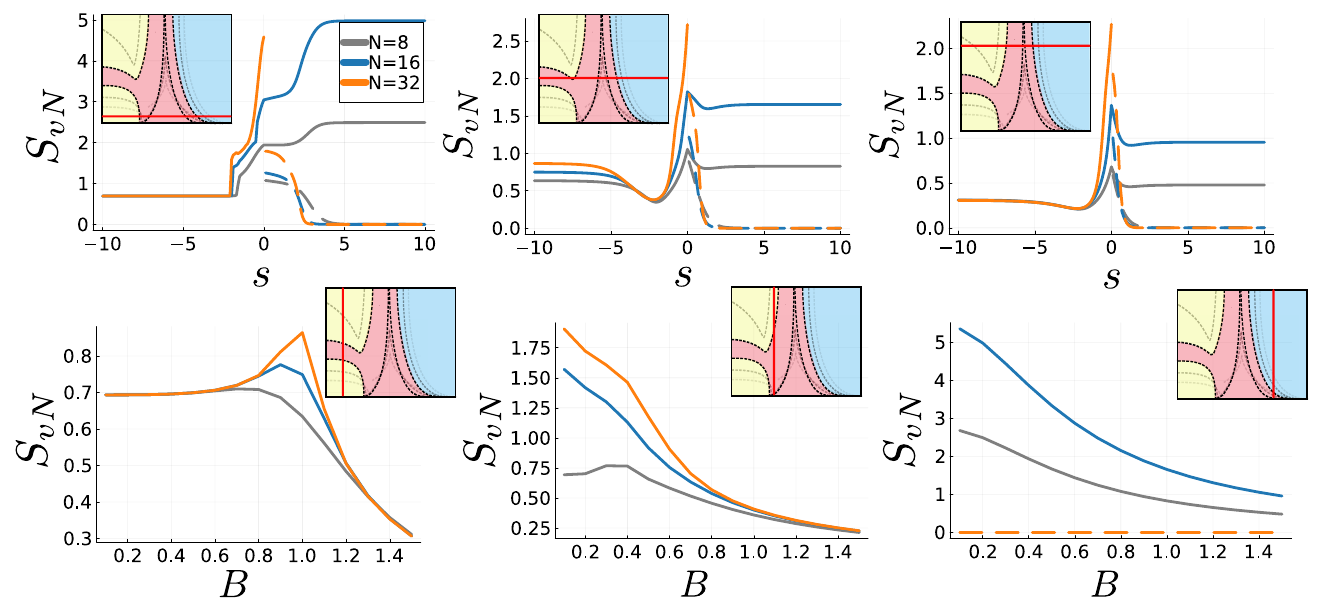}
    \caption{Half-chain entanglement entropy $S_{vN}$ as a function of coupling parameter $s$ and transverse field strength $B$ captures the phase structure of the quantum model. Six representative slices are shown at $B/J = 0.2,1.0,1.5$ (top row) and $s = -5,-1.8,5$ (bottom row), capturing distinct behaviors across the diagram. These slices correspond to key regimes highlighted in the schematic phase diagram of Fig.~\ref{fig:1}(c), including regions of area-law, volume-law, and intermediate entanglement scaling, as well as the distinct ground-state structure (labeled (i)-(iii)). Solid lines reflect Euclidean site ordering in $S_{vN}$ calculation. Dashed lines represent Monna-mapped site ordering.}
    \label{fig:svn}
\end{figure}

The quantum model with $B\neq 0$ can not be analysed using the MCMC methods from above. The Density Matrix Renormalization Group (DMRG) and matrix product states (MPS) enable the study of ground-state properties for significantly larger 1D quantum systems. A key limitation of DMRG is its poor convergence with periodic boundary conditions (PBC), due to the long-range bond between the first and last sites. While Phase 1 remains accessible, being short-range and local, exploring other phases becomes increasingly challenging as we increase $s$. However, for $s>0$, we can exploit our alternative notion of locality by applying a Monna map to the MPS site ordering. This reorders sites according to their $2$-adic representation, placing strongly correlated (distant) sites adjacent in the MPS structure. As a result, long-range correlations are encoded locally, reducing the reliance on long bonds and improving convergence. This approach aligns the tensor network structure with the system's intrinsic ultrametric geometry, enabling more efficient simulations beyond Phase 1. Convergence remains difficult in the intermediate $s$-regime so we use system-sizes $N=8,16,32$ and use finite-size scaling to extract estimates of the thermodynamic limit.

\paragraph{Weak quantum fluctuations $B\ll1$ and finite-size effects.} As shown in Fig.~\ref{fig:lowB}, for small $B$ and finite $N$, elements of the classical phase diagram survive. The behavior of the entanglement entropy in this region can be fully described by a comparison to the classical phase diagram. Note that five regions have been highlighted in this figure, compared to the four regions in Fig.~\ref{fig:2}, as we are discussing the finite size regions rather than the thermodynamic phases. Region 1 (nearest neighbor region) is visible in the entanglement entropy analysis, clearly characterized by the NN AFM ground-state which has a constant entanglement entropy $S_{vN} = \log(2)$ as expected for the entropy of a maximally mixed 2-state subsystem with ground-state $\ket{\psi} = \frac{1}{\sqrt{2}} \left( \ket{\uparrow \downarrow \uparrow \dots} + \ket{\downarrow \uparrow \downarrow \dots} \right)$. The entropy stays constant even as the energy gap is closing in this region as the first excited band is not the single spin flip. In Region 2, the entropy sharply transitions and you can see evidence of the finite-size effects at this system size. For larger system sizes where this region is characterized by a zero energy gap, the entropy plateaus to a constant here. The ground-state in this region is a combination of the classical ground-states and the excited states, as they are easy to access in a gapless region. In Region 3, the energy gap opens. It becomes harder to access excited states and they will not contribute significantly to the ground-state. This allows for more structured correlations to appear in the ground-state from the translational invariance and less ``noise" from the excitations, so the entanglement entropy grows. The entanglement entropy then plateaus on crossing to $s>0$ reflecting the energy gap plateau in the classical phase. In Region 4, the entanglement entropy grows, as the manifold of $2^{N/2}$ states that satisfy all furthest neighbor bonds collapse onto the classical ground-state in our energy gap plot of Fig.~\ref{fig:lowB}. In Region 5, all states that satisfy FN bonds contribute equally to the ground-state, the entanglement entropy plateaus at a value $\log(2^{N/2}) \approx 5.55$ for $N=16$. 

\paragraph{Larger system sizes.} The low $B$ behavior described above holds as we scale system size but there is one crucial change. The transverse field merges the two classical phases 3 and 4 (finite regions 4 and 5 in Fig.~\ref{fig:lowB}) into a single quantum transition region. This stems from the fact that between those two classical critical points there was never a distinct ordered phase, just a collapsing manifold, therefore, the quantum perturbations simply blur that interval into one continuous crossover. The quantum critical line that leaves the positive-s axis originates near the onset of manifold formation, because that’s where the first true level crossing and change of the ground-state character occur. The second classical point, where the manifold finishes collapsing and a new gap reopens, doesn’t yield a separate quantum line. Representative cuts through Fig.~\ref{fig:svn}(b) illustrates the behavior of the bipartite entanglement entropy for increasing system sizes, with the colored regions in Fig.~\ref{fig:1}(c) directly reflected in the entanglement scaling profiles here. The post-Monna mapping area-law behavior at $s>0$ is shown as dashed lines. A comparison of system sizes $N=8,16,32$ reveals the scaling of the critical points against quantum fluctuations. The only classical phases that survive in an infinite system are the NN AFM phase and a critical line at $s=0$. There is a chance that other phase lines exist in our diagram that can be captured using different order parameters. For example, in our gapless region between $s_c=-2$ and $s_c=0$ it would be interesting to explore the existence of possible spin liquids or other commensurate-incommensurate transitions that are possible in systems with frustration and competing AFM couplings.

\section{Finite-size scaling and uncertainties}

We determine an estimate for the thermodynamic critical point $s_c$, dynamical exponent $z$, and correlation-length exponent $\nu$ using pairwise finite-size scaling on system-size pairs $(N,2N)$, followed by an extrapolation to the thermodynamic limit. Throughout we include at least three pairs so that the extrapolation fit has at least one degree of freedom. This ensures our standard errors are meaningful, and naturally led us to choose $B$ large enough ($B=0.8$) for stable DMRG convergence up to $N= 256$.

\begin{figure}[h!]
    \centering
    \includegraphics[width=0.7\linewidth]{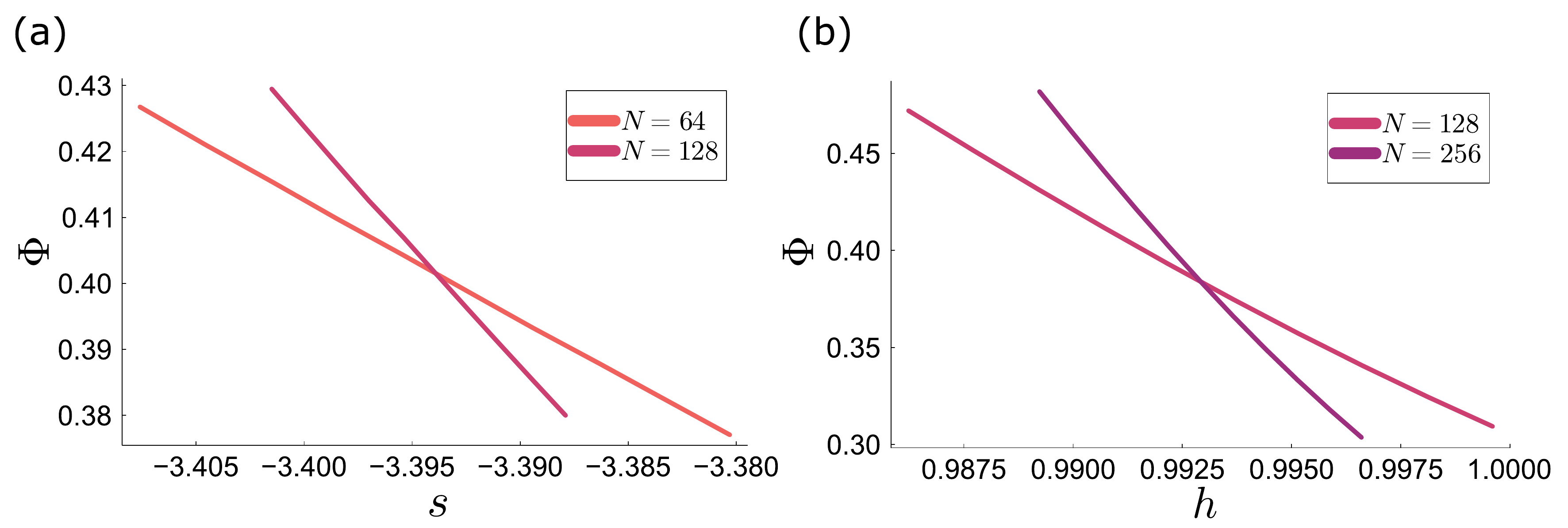}
    \caption{ (a) Crossing of the structure factor $\Phi_N(s)=\xi_N(s)/N$ for the PWR2 model using system sizes $(64,128)$, yielding a critical-point estimate $s_{c\times}\approx -3.3939$. (b) Corresponding crossings for the Rydberg tambourine model using sizes $(128,256)$, giving $h_{c\times}\approx 0.9935$. In both cases, $\Phi_N$ uses the second-moment correlation length extracted from the connected structure factor [Eq.~\eqref{eq:xi-second-moment}].}
    \label{fig:structure}
\end{figure}

\medskip
\noindent\textit{Locating $s_c$ from the structure factor.}\; We locate the critical point using the size-independent crossing of $\Phi_N(s)\equiv \xi_N(s)/N$, where $\xi_N$ is the second-moment correlation length extracted from the static connected structure factor $S(q)$,
\begin{equation}
S(q)=\frac{1}{N}\sum_{i,j} e^{iq(i-j)}\!\left[ \langle S^z_i S^z_j\rangle - \langle S^z_i\rangle\langle S^z_j\rangle \right].
\end{equation}
 A symmetric second-moment estimator is \cite{Sandvik_2010},
\begin{equation}
\label{eq:xi-second-moment}
\xi(q_0)=\frac{1}{2\sin(\delta/2)}
\sqrt{
\frac{S(q_0)}{\tfrac12\,[S(q_0+\delta)+S(q_0-\delta)]}-1
}\, ,
\end{equation}

where $\delta=2\pi/N$. We take $q_0=\pi$ as we have antiferromagnetic couplings. The crossing of $\Phi_N(s)$ and $\Phi_{2N}(s)$ then defines a pairwise estimate $s_c(N,2N)$. The resulting crossings, shown in Fig.~\ref{fig:structure}, give $s_{c\times}\approx -3.3939$ (PWR2) and $h_{c\times}\approx 0.9935$ (Rydberg). This criterion avoids the circularity of gap-based crossings where extraction of the critical point and dynamic exponent $z$ are dependent. 

\medskip
\noindent\textit{Estimating $\nu$ from structure--factor slopes.}\; 
We exploit the formula $\Phi_N(s)=f[(s-s_c)N^{\frac{1}{\nu}}]$ at the pairwise crossing $s_c(N,2N)$ of $\Phi_N(s)$ and $\Phi_{2N}(s)$. We estimate $\nu$ from the ratio of local slopes,
\begin{equation}
\nu_{N,2N}
=\log_2\!\big[(d\Phi/ds)_{2N}/(d\Phi/ds)_{N}\big]\,.
\end{equation}
The derivatives $(d\Phi/ds)_N$ are obtained by a symmetric fit in a \emph{scaled} window $\lvert x\rvert\le x_{\rm win}$ with $x=(s-s_c)N^{1/\nu_{\rm guess}}$ where we take $x_{\rm win} \in [0.25,0.5]$ and $\nu_{\rm guess} = 1.0$. This keeps the two sides comparable and suppresses bias from asymmetric sampling around $s_c$. Error bars for $(d\Phi/ds)_N$ come from the standard error of the local linear regression in that window $\sigma_{(d\Phi/ds)}$; these are then propagated to $\nu_{N,2N}$ via the propagation of uncertainty formula,
\begin{equation}
\sigma_\nu^2
\simeq \frac{1}{\ln(2)^2}
\left[\left(\frac{\sigma_{(d\Phi/ds)_{2N}}}{(d\Phi/ds)_{2N}}\right)^2
+\left(\frac{\sigma_{(d\Phi/ds)_{N}}}{(d\Phi/ds)_{N}}\right)^2\right].
\end{equation}

\medskip
\noindent\textit{Estimating $z$ from energy--gap ratios.}\; 
At the same $s_c(N,2N)$ we compute the lowest gap $\Delta_N$ and estimate $z$ from the ratio,
\begin{equation}
z_{N,2N}
=-\log_2\!\big[\Delta_{2N}(s_c)/\Delta_{N}(s_c)\big]\,.
\end{equation}
In our data the uncertainties of $\Delta_N$ and $s_c$ are negligible compared to finite--size systematics, so we assign a small statistical floor to $\sigma_z$ and let the systematic analysis (see below) set the dominant error bar.

\medskip
\noindent\textit{Thermodynamic extrapolation.}\; 
For each exponent $y\in\{\nu,z\}$ we assemble the pairwise sequence $y_{N,2N}$ versus the effective size $N_{\rm eff}=\sqrt{N\,2N}$ and perform \emph{weighted} least--squares fits with the ans\"atz for the leading corrections:
$
\quad y_{N,2N}=y_\infty+a\,N_{\rm eff}^{-\omega}\;\;(\omega\;\text{free}).
$
Weights are $w_i=1/\sigma_{y,i}^2$, where $\sigma_{y,i}$ are the pairwise standard errors (for $z$ we use the conservative floor noted above). The quoted statistical uncertainty on $y_\infty$ is the standard error of the fitted intercept from the weighted regression. 

\medskip
\noindent\textit{Systematic uncertainties and final error bars.}\; 
Finite--size systematics are quantified by a \emph{leave--one--out} sensitivity,
$
\mathrm{\sigma}_{\mathrm{sys}}
=\max\!\Big(\big|y_\infty^{\mathrm{all}}-y_\infty^{(-\mathrm{small})}\big|,\;
             \big|y_\infty^{\mathrm{all}}-y_\infty^{(-\mathrm{large})}\big|\Big).
$
The final quoted error is the quadrature sum
$
\sigma_{\mathrm{final}}
=\sqrt{\sigma_{\mathrm{stat}}^2+\mathrm{\sigma}_{\mathrm{sys}}^2}.
$
Fig.~\ref{fig:crit}(a) \& (b) shows the thermodynamic extraction and best fit curve for $y_{\infty}$ for both models.

\medskip
\noindent\textit{Central charge.}\; For $c$ we use the largest available system size and fit the Calabrese-Cardy form for PBC \cite{Pasquale_Calabrese_2004,Cardy_2010},
\[
S_{\rm vN}(\ell)=\frac{c}{3}\,\ln\!\Big[\frac{N}{\pi}\sin\!\Big(\frac{\pi\ell}{N}\Big)\Big]+\mathcal{O}(1)\,,
\]
with $\ell=1,\dots,L/2$. Defining $x_\ell=\ln\!\big[\frac{N}{\pi}\sin(\pi\ell/N)\big]$, we perform an ordinary least-squares fit of $S_{\rm vN}(\ell)$ versus $x_\ell$ on a central window (e.g.\ $\ell\in[\log_2(N), \;N-\log_2(N)]$) to suppress subleading oscillations. We also refine $c$ by allowing it to vary within the uncertainty window of the corresponding critical-point estimate $s_c$. Fig.~\ref{fig:crit}(c) \& (d) shows the fitted slope $b$ and its standard error $\sigma_b$ give
\[
c=3\,b,\qquad \delta c=3\,\sigma_b.
\]

\section{Mapping of $PWR2$ control parameter $s$ and Dual-Species Implementation}

Our proposed experiment promotes the \emph{graph} geometry of the power-of-two (PWR2) couplings to the \emph{real-space} geometry of a Rydberg ring by displacing every second atom out of plane as illustrated in Fig.~\ref{fig:1}(d). Our control parameter $s$ in the PWR2 model must map directly onto the control parameter $h$ in the tambourine Rydberg model, since both $s$ and $h$ tune the vertical displacement between the two rings (red atoms and black atoms in \ref{fig:1}(d) in the tambourine geometry). 

\begin{figure}[h!]
    \centering
    \includegraphics[width=0.6\linewidth]{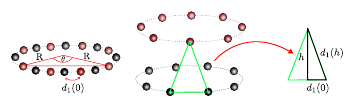}
    \caption{Nearest-neighbour (NN) distance $d_{1}(h)$ in the “tambourine’’ ring geometry as a function of the staggered vertical displacement amplitude $h$ (in units of the in-plane spacing). We fix the undistorted NN spacing to unity, so $d_{1}(0)=1$ and $d_{1}(h)=\sqrt{1+h^{2}}$. Throughout, all separations are renormalized by $d_{1}(h)$, which pins the $d=1$ bond while the remaining distances $d_{j}(h)$ (and corresponding couplings $V_{j}(h)\propto d_{j}(h)^{-6}$) vary with $h$.}
    \label{fig:mapping}
\end{figure}

\paragraph{Mapping between Rydberg displacement $z$ and power--law exponent $s$.}
We begin by considering a ring of $N$ atoms before any vertical displacement, as shown in Fig.~\ref{fig:mapping}(left). The radius of the ring $R=\frac{1}{2\sin(\pi/N)}$. The in-plane chord distances between sites separated by $k$ steps are then,
\begin{equation}
    d_k(h=0) = 2R\sin{\left(\frac{\pi}{N}\right)}.
\end{equation}
If we normalize in terms of nearest-neighbour spacing,
\begin{equation}
\tilde{d}_k(h=0) = \frac{\sin(\pi k / N)}{\sin(\pi / N)}.
\end{equation}
When we introduce an alternating vertical displacement, with every second site shifted out of plane by a height $h$, 
the nearest-neighbour distance becomes,
\begin{equation}
d_1(h) = \sqrt{1 + h^2}.
\end{equation}
as evident in Fig.~\ref{fig:mapping}(right). As we tune $h$, all other distances between atoms are renormalized relative to the changing nearest-neighbour distance,
\begin{equation}
\tilde{d}_k(h) = \frac{d_k(h)}{d_1(h)}.
\end{equation}
This normalization ensures that the nearest-neighbour bond strength remains fixed as the geometry is deformed.
For even $k$, the two sites lie on the same ring, so $d_k(h) = d_k(0)$, while for odd $k$ the two sites reside on different rings, giving
\begin{equation}
d_k(h) =
\begin{cases}
\sqrt{d_k(0)^2 + h^2}, & k~\text{odd},\\
d_k(0), & k~\text{even}.
\end{cases}
\end{equation}

These expressions describe how the interatomic distances change continuously with $h$ in the ``tambourine'' geometry.  
Since the van der Waals couplings scale as $V_k(h) \propto d_k(h)^{-6}$, the change in bond strengths can be written explicitly as
\begin{equation}
V_k(h) = \tilde{d}_k(h)^{-6}.
\end{equation}

To complete the mapping from $h$ to $s$, we identify that $V_k(h)=J_k(s) = k^{\,s}$.
This provides a direct mapping between geometry and exponent:
\begin{equation}
s_k(h) = \log_k V_k(h) = \frac{\ln V_k(h)}{\ln k}.
\end{equation}

At the PWR2 critical point, the couplings at distances $d\in\{1,2,4,8\}$ control the physics. We know this as truncating the PWR2 graph shifts the apparent critical point away from $s_c=-2$ (e.g. drifting to $\simeq-1.6$ for including only $d=1,2,4$). However, including bonds up to $d=8$ restores the collapse at $s_c=-2$ and so captures the physics of the model in this $s$-regime. Because $d_1(h)$ is fixed by renormalization, $V_1(h) = 1$, and we can define an effective exponent as a combination of the $d=2,4,8$ contributions:
\begin{equation}
s(h) = \tfrac{1}{3}\!\left[ s_2(h) + s_4(h) + s_8(h) \right]
= \tfrac{1}{3}\!\left[ \log_2 V_2(h) + \log_4 V_4(h) + \log_8 V_8(h)\right],
\end{equation}
which ensures dominant couplings are simultaneously well approximated by the same effective power-law exponent. With this mapping, the tambourine Rydberg model’s classical critical point doesn’t land at $s_c=-2$; instead it shifts toward $s_c\approx -3$. The reason is that the Rydberg implementation isn’t sparse--non-PWR2 couplings are present--so both the location of the transition and the post-critical ground-state are altered (the latter no longer matches the recursive PWR2 state). To remedy this, we introduce a dual-species scheme that selectively strengthens PWR2 distances and suppresses non-PWR2 bonds, which restores the expected critical point and the intended PWR2 ground-state.\\

\paragraph{Dual-species implementation.} A complementary improvement is to engineer the effective couplings by using two different species arranged to strengthen PWR2 distances and dampen non-PWR2 distances. Dual-species configurations have recently been proposed and realised to control inter- and intra-species interaction strengths in Rydberg arrays \cite{Kevin_Singh_2022,Kevin_Singh_2023,Beterov_2015, Samboy_2017,Ireland_Paul_2024}, offering a natural framework for tailoring specific coupling hierarchies.

\begin{figure}
    \centering
    \includegraphics[width=0.35\linewidth]{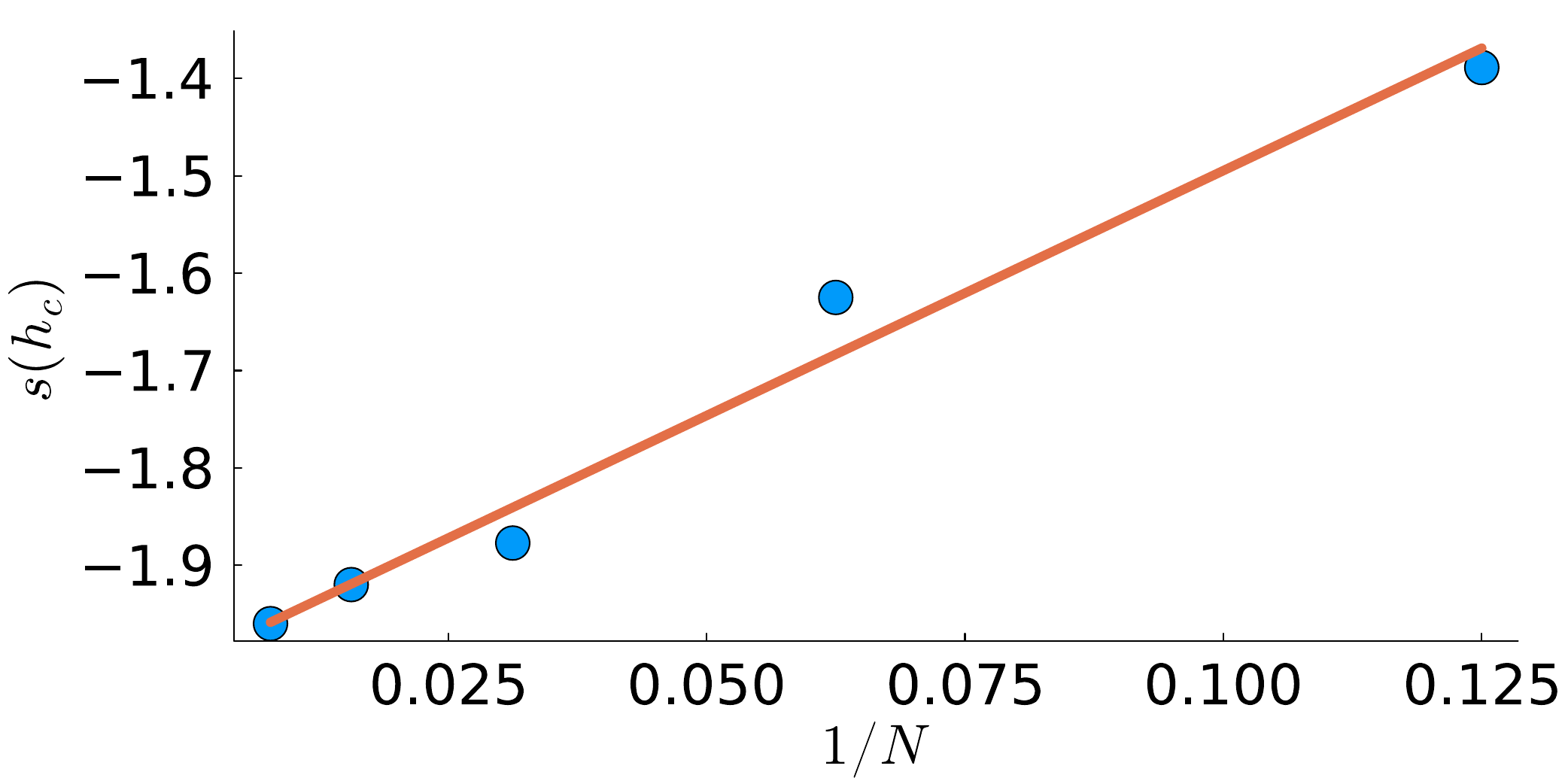}
    \caption{Finite-size scaling of the mapped exponent at the Rydberg critical displacement. Plotted is \(s(h_c;N)\equiv s_c(N)\) versus \(1/N\) for $N=8,16,32,64,128$; the solid line is a fit linear in \(1/N\). Extrapolating to \(1/N\to 0\) yields \(s(h_c) = -1.998(7)\), consistent with the PWR2 critical point \(s_c=-2\). Agreement is enabled by the dual-species implementation, which suppresses non-PWR2 couplings and restores the PWR2 critical point.}
    \label{fig:h_c}
\end{figure}

With this approach, we exploit the Hamiltonian,

$$
\begin{aligned}
H_{\mathrm{tot}}
&= \sum_{i\in\mathcal{A}} \frac{B}{2}\,\sigma_i^{x}
 \;+\; \sum_{i\in\mathcal{B}} \frac{B}{2}\,\sigma_i^{x} \\
&\quad+\; \sum_{\substack{i<j\\ i,j\in\mathcal{A}}} V^{\mathrm{a\!-\!a}}_{ij}\,S_i^z S_j^z
 \;+\; \sum_{\substack{i<j\\ i,j\in\mathcal{B}}} V^{\mathrm{b\!-\!b}}_{ij}\,S_i^z S_j^z
 \;+\; \sum_{\substack{i\in\mathcal{A}\\ j\in\mathcal{B}}} V^{\mathrm{a\!-\!b}}_{ij}\,S_i^z S_j^z \,,
\end{aligned}
$$

$$
V^{\mathrm{A\!-\!A}}=V^{\mathrm{B\!-\!B}} = \frac{1}{r_{i,j}^6},\qquad V^{\mathrm{A\!-\!B}} = \frac{1}{r_{i,j}^3}, \qquad
\mathcal{A}=\text{Species (a)},\quad
\mathcal{B}=\text{Species (b)}.
$$
We find that the optimal two-species arrangement is the recursive pattern $[a,b,b,a,b,a,a,b,\ldots]$, guided by the same construction as the recursive PWR2 ground-state. This pattern dampens non-PWR2 couplings and strengthens the $d=2^k$ bonds, thereby shifting the critical point back toward $s_c\approx-2$ (Fig.~\ref{fig:h_c}). With this implementation the post-critical ground-state recovers the recursive PWR2 ground-state, showing that we have sufficiently dampened the unwanted terms. In this way we address the two main concerns—(i) the mapping $h_c \to s_c$ and (ii) the presence of non-PWR2 bonds—while preserving the observed Ising critical behavior.

\end{document}